\documentclass{moriond}

\usepackage{amsmath}
\usepackage{amssymb}

\def\be{\begin{equation}}
\def\ee{\end{equation}}
\def\bea{\begin{eqnarray}}
\def\eea{\end{eqnarray}}

\newcommand{\Photo}{}

\begin{document}

\hfill FLAVOUR(267104)-ERC-72

\vspace*{4cm}

\title{UPDATED $\boldsymbol{B_q \to \bar\ell\ell}$ IN THE STANDARD MODEL AT HIGHER ORDERS}

\author{CHRISTOPH BOBETH}

\address{  
  Technische Universit\"at M\"unchen, \\
  Institute for Advanced Study, \\
  D-85748 Garching,
  Germany
}

\maketitle\abstracts{ The latest standard model predictions of rare leptonic
  $B_q \to \bar\ell\ell$ decays are reviewed including recent results of
  next-to-leading electroweak and next-to-next-to-leading QCD corrections.}

%
%
%

\section{Introduction}

The rare leptonic $B_qÊ\to \bar\ell\ell$ decays of neutral $B$-mesons -- with $q
= d, s$ and $\ell = e, \mu, \tau$ -- provide interesting tests of
flavor-changing neutral-current (FCNC) transitions mediated by $b\to q\,
\bar\ell\ell$ at the parton level as predicted by the standard model
(SM). Besides the loop-suppression of FCNC decays in the SM, the SM contribution
is in addition helicity suppressed, which leads the very rare branching ratios
$\sim {\cal O}(10^{-9} - 10^{-10})$ for $\ell = \mu$, and offers tests of
non-standard effects due to scalar and pseudo-scalar interactions.


Utilising their full data sets of $3\,$fb$^{-1}$ and $25\,$fb$^{-1}$ by the year
2013, the two experiments LHCb~\cite{Aaij:2013aka} and
CMS~\cite{Chatrchyan:2013bka} succeeded to observe the CP-averaged
time-integrated branching ratio~\cite{DeBruyn:2012wk} of the $B_s \to
\bar\mu\mu$ channel with a statistical significance of $4.0\,\sigma$ and
$4.3\,\sigma$, respectively. The evidence of the measurement of the other muonic
channel, $B_d\to \bar\mu\mu$, is about $2.0\,\sigma$ in both experiments and
correlated with $B_s \to \bar\mu\mu$. The combination of the LHCb and CMS
measurements \cite{LHCb-CMS:2013:EPSHEP} yields
\begin{equation}
  \label{eq:exp:Br}
\begin{aligned}
  \overline{{\cal B}}[B_s \to \bar\mu\mu]_{\rm exp.} &
  = (2.9 \pm 0.7) \times 10^{-9}\,,
\\[0.1cm]
  \overline{{\cal B}}[B_d \to \bar\mu\mu]_{\rm exp.} &
  = (3.6^{+1.6}_{-1.4}) \times 10^{-10}\,,
\end{aligned}
\end{equation}
where the uncertainties are currently dominated by the statistical error, but
include systematic sources, as well. As a consequence of the combination of LHCb
and CMS measurements one can consider the decay $B_s \to \bar\mu\mu$ as observed
(i.e. $> 5\, \sigma$) while the yield of $B_d \to \bar\mu\mu$ is not
statistically significant yet (i.e. $< 3\, \sigma$).


The future experimental prospects are rather promising for $B_s \to \bar\mu\mu$.
For example LHCb is expected \cite{Bediaga:2012py} to reach an accuracy of
$0.5\times 10^{-9}$ after Run 2 of LHC in the years 2015-2017 and even
$0.15\times 10^{-9}$ after the detector upgrade with a data set of 50
fb${}^{-1}$. The latter absolute error corresponds to about $5\,\%$ relative
error given the current central value of the measured branching ratio. Further,
the large data set allows to measure the ratio ${\cal R}_\mu \equiv
\overline{{\cal B}} [B_d \to \bar\mu\mu] / \overline{{\cal B}} [B_s \to
\bar\mu\mu]$ with a relative error of $35\,\%$.  A recent comprehensive study
\cite{CMS:2013:prospects} of the reach of CMS expects that with the current
detector Run 2 of LHC will deliver about
100~fb${}^{-1}$, which enables CMS to measure $\overline{{\cal B}}[B_s \to
\bar\mu\mu]$ with a significance $> 10\, \sigma$ and a relative error of
$15\,\%$ and for ${\cal R}_\mu$ about $70\,\%$. After Run 3 of LHC in the years
2019-2021, these relative errors become $12\,\%$ and $50\,\%$, respectively,
based on a data set of about 300~fb${}^{-1}$, and the statistical significance
for the decay $B_d \to \bar\mu\mu$ might reach the $3\,\sigma$ level. In the far
future, the potential luminosity upgrade of the LHC machine to HL-LHC (starting
with the year 2023) will deliver about 3000~fb${}^{-1}$, however, without
substantial reduction of currently assumed systematic errors, the relative error
of $\overline{{\cal B}}[B_s \to \bar\mu\mu]$ will remain around $12\,\%$,
whereas $\overline{{\cal B}} [B_d \to \bar\mu\mu]$ can be discovered (i.e. $>
5\,\sigma$) and a relative error of $18\,\%$ can be reached. The latter would
imply about $21\,\%$ relative error in ${\cal R}_\mu$.


In view of future experimental uncertainties below $10\,\%$, the theoretical
predictions have been updated in several respects over the last two years. These
efforts were further motivated by the steady progress of lattice determinations
of the $B_q$-decay constant that constituted in the past the major source of
uncertainty and has been reduced nowadays to the level of other parametric
uncertainties.  First, we will give a short introduction of the theoretical
treatment of $B_q \to \bar\ell\ell$ decays including higher order electroweak
(EW) and QCD corrections in section~\ref{sec:theory}. Then we will discuss in
section~\ref{sec:numeric} the latest SM predictions and present a detailed
uncertainty budget, which can bear comparison with experimental prospects.

%
%
%

\section{Theory of $\boldsymbol{B_q \to \bar\ell\ell}$
  \label{sec:theory}
}

Rare $\Delta B = 1$ FCNC decays can be conveniently described by an effective
theory of the electroweak interactions after the application of an
operator product expansion (OPE). The OPE corresponds to an expansion in external
momenta of the considered processes that are of the
order of the bottom-quark mass, $m_b \sim 5$ GeV, which are much smaller than the
internal heavy $W$-boson mass $m_W \sim 80$ GeV. The effective Lagrangian takes
a systematic expansion
\begin{align}
  \label{eq:eff:Lag}
  {\cal L}_{\rm eff} & 
  = {\cal L}_{\rm QCD \times QED} (u,d,s,c,b, \,e,\mu,\tau)
  + {\cal L}_{\rm dim = 6}
  + {\cal L}_{\rm dim = 8}
  + \ldots
\end{align}
where the first term describes the QCD and QED gauge interactions of $\rm dim =
4$ of the light ($N_f = 5$) quarks and leptons. The second term represents the
leading effect of flavor-changing operators $O_i$ of $\rm dim = 6$ that mediate
$|\Delta B|Ê= |\Delta Q| = 1$ ($Q = D, S$) processes
\begin{align}
  \label{eq:dim-6:eff:Lag}
  {\cal L}_{\rm dim = 6} & 
    = {\cal N}_{\rm eff}^{\,q}\, \sum_i C_i(\mu_b) \, O_i + \mbox{h.c.} \,, &
  {\cal N}_{\rm eff}^{\,q} & 
    = \frac{G_F^2 m_W^2}{\pi^2} V_{tb}^{} V_{tq}^\ast \,,
\end{align}
whereas the higher dimensional operators ($\rm dim > 6$) are suppressed by
$(m_b/m_W)^2 \sim 0.3 \,\%$. Each operator has an effective coupling $C_i$
(Wilson coefficient), which is determined in the matching at a so-called
matching scale of order $m_W$ and has been evolved with the aid of the
renormalization group (RG) equation of the effective theory to the low-energy
scale $\mu_b \sim m_b$. This procedure provides a systematic framework to
account for the large logarithmic contributions $\sim \alpha_s^n
\ln^n(m_W/\mu_b)$ from radiative corrections to all orders in the QCD coupling
$\alpha_s$. Nowadays, the $C_i(\mu_b)$
relevant for $b\to q\,\bar\ell\ell$ transitions are known to rather high orders
including the RG evolution \cite{Bobeth:2003at,Huber:2005ig}. The choice of the
normalization factor \cite{Misiak:2011bf} ${\cal N}_{\rm eff}^{\,q}$ provides better
convergence properties with regard to higher order EW corrections. It contains
besides Fermi's constant, $G_F$, the product of elements of the
Cabibbo-Kobayashi-Maskawa (CKM) quark-mixing matrix.

Concerning $B_q\to\bar\ell\ell$, at leading order (LO) in EW/QED interactions,
one single operator
\begin{align}
  O_{10} & 
    = \big(\bar{q} \gamma_\mu P_L b\big) 
      \big(\bar\ell \gamma^\mu \gamma_5 \ell\big) \,, &
  P_L = \frac{1-\gamma_5}{2}
\end{align}
is relevant. Since $O_{10}$ is a conserved current under QCD to this order in
EW/QED interactions, it's Wilson coefficient \footnote{Here we use the
  convention $C_{10} = - 2\, C_A$ compared to
  \cite{Hermann:2013kca,Bobeth:2013uxa} and $C_{10} = \widetilde{c}_{10}$ to
  \cite{Bobeth:2013tba}. It differs by a factor of sine squared of the weak
  mixing angle to $c_{10}$ of \cite{Bobeth:2003at,Huber:2005ig}: $C_{10} = s_W^2
  c_{10}$ at LO in EW interactions.}  does not evolve and does not depend on
$\mu_b$. The LO contribution~\cite{Inami:1980fz} of $C_{10}$ has as strong
dependence on the renormalization scheme of the top-quark mass, which is
cancelled by the next-to-leading
\cite{Buchalla:1992zm,Buchalla:1993bv,Misiak:1999yg,Buchalla:1998ba} (NLO) and
next-to-next-to-leading \cite{Hermann:2013kca} (NNLO) order QCD corrections.
The inclusion of NNLO corrections reduce the related uncertainties from
$1.8\,\%$ at NLO to less than $0.2\,\%$ at the level of the branching ratio
\cite{Hermann:2013kca}.

With this remarkable control of short-distance QCD corrections, also higher
order EW/QED corrections have to be considered. NLO EW corrections to the
matching of $C_{10}$ had been derived in the large top-quark mass limit
\cite{Buchalla:1997kz} and used to point out that EW renormalization scheme
dependences \cite{Buras:2012ru} of the branching ratio are a sizeable source of
$\sim 7\,\%$ to $5\,\%$ uncertainty.  Recently, the complete NLO EW matching
corrections to $C_{10}$ have been calculated \cite{Bobeth:2013tba} employing
several different choices of renormalization schemes of the relevant
parameters. The best convergence properties, i.e., small NLO corrections
compared to the LO contribution, were obtained in the scheme that eliminates the
ratio $\alpha_e/s_W^2$ in favor of $G_F$ -- as done with the choice of ${\cal
  N}_{\rm eff}^{\,q}$ -- and in the scheme where both quantities entering the ratio
$\alpha_e/s_W^2$ are renormalized in the $\overline{\rm MS}$
scheme~\cite{Brod:2010hi}. Due to the removal of EW scheme dependences at NLO,
the previous $\sim 7\,\%$ uncertainty of the branching ratio at LO was reduced
to about $0.6\,\%$ at NLO. Furthermore, at NLO in QED, operator mixing of
$O_{10}$ with other operators in (\ref{eq:dim-6:eff:Lag}) leads to a non-trivial
evolution of $C_{10}$ and gives rise to $\mu_b$-dependence. The necessary
ingredients for a consistent RG evolution had been provided before
\cite{Bobeth:2003at,Huber:2005ig}, with explicit results given in
\cite{Bobeth:2013tba}.

At LO in QED, the matrix element of $B_q\to \bar\ell\ell$ is easily derived
from (\ref{eq:dim-6:eff:Lag}) by restricting the sum to $i = 10$ and neglecting
QED interactions below the scale $\mu_b$, i.e., replacing ${\cal L}_{\rm QCD \times QED}
\to {\cal L}_{\rm QCD}$
\begin{equation}
\begin{aligned}
  \label{eq:LOQED:ME}
  \frac{i {\cal M}_{\rm LO\, QED}}{{\cal N}_{\rm eff}^{\,q}\,ÊC_{10}(\mu_b)} &
  = \, \big\langle \bar\ell\ell \big| O_{10} \big| B_q \big\rangle
  = \big\langle \bar\ell\ell \big| \bar\ell \gamma^\mu \gamma_5 \ell \big| 0 \big\rangle
      \big\langle 0 \big|\bar{q} \gamma_\mu P_L b \big| B_q \big\rangle
  =  m_\ell (\bar{u}_\ell \gamma_5 v_\ell) f_{B_q}\,.
\end{aligned}
\end{equation}
This allows to factorize the currents present in $O_{10}$ and the $B_q$-decay 
constant $f_{B_q}$ is determined in lattice calculations \cite{Aoki:2013ldr}.
Here we include NLO EW matching corrections in $C_{10}$ as they are complete 
and do not cause $\mu_b$ dependence. The latter 
is entirely due to photonic corrections within the effective theory and 
will be cancelled by NLO QED corrections to the matrix element. The scale
variation of $\mu_b$ from $m_b/2$ to $2 m_b$ leads to a variation of 
the branching ratio of $0.3\,\%$, which corresponds to a typical size of
a ${\cal O}(\alpha_e)$ correction. The actual calculation of the lacking
virtual NLO QED corrections to the matrix elements seems very
challenging bearing in mind that QED corrections prevent a simple
factorization of the matrix element as in (\ref{eq:LOQED:ME}) at LO.

The average time-integrated branching ratio of $B_q \to \bar\ell\ell$
can be obtained to a very high accuracy in the SM as $\overline{{\cal B}} 
\big[ B_q \to \bar\ell\ell \big] = \Gamma \big[ B_q \to \bar\ell\ell \big]/
\,\Gamma_H^q$ where $\Gamma_H^q$ denotes the heavier mass-eigenstate's
total width \cite{DeBruyn:2012wk}. With (\ref{eq:dim-6:eff:Lag}) and
(\ref{eq:LOQED:ME}), it takes the form
\begin{align}
  \label{eq:BR}
  \overline{{\cal B}}\big[ B_q \to \bar\ell\ell \big]
  = \frac{|{\cal N}_{\rm eff}^{\,q}|^2\, m_{B_q}^3\, f_{B_q}^2}{8 \pi\, \Gamma_H^q}
    \left(\frac{m_\ell}{m_{B_q}}\right)^2 \sqrt{1 - \frac{4 m_\ell^2}{m_{B_q}^2}}
    \,\, \Big|C_{10}(\mu_b)\Big|^2
    + {\cal O}(\alpha_e)
\end{align}
exhibiting the helicity-suppression factor $m_\ell/m_{B_q}$ of the lepton and
$B_q$-meson masses. The ${\cal O}(\alpha_e)$ term represents now the previously
discussed virtual NLO QED corrections as well as soft photon bremsstrahlung. 
The latter might receive large enhancements in the presence of kinematical cuts
in the experimental analysis, or initial-state radiation of photons, which may lift the
helicity suppression. The intial-state radiation is infrared safe due to electrically
neutral $B_q$ meson and it's interference with final-state radiation is helicity
suppressed. Moreover, it is strongly phase space suppressed \cite{Aditya:2012im} 
within the signal window in the dilepton invariant mass chosen by LHCb
\cite{Aaij:2013aka} and CMS \cite{Chatrchyan:2013bka}, which allows to ignore it
in the experimental analysis and discard it on the theoretical side
by definition. Concerning photon bremsstrahlung from the final-state leptons,
LHCb \cite{Aaij:2013aka} and CMS \cite{Chatrchyan:2013bka} apply PHOTOS
\cite{Golonka:2005pn} to extrapolate beyond cuts in the dilepton invariant mass,
such that soft QED logarithms cancel out and the experimental quantity (\ref{eq:exp:Br})
is equivalent to (\ref{eq:BR}) up to terms that do not receive an extra enhancement 
\cite{Buras:2012ru}. 
 
%
%
%

\section{Updated prediction of $\boldsymbol{B_q \to \bar\ell\ell}$
  \label{sec:numeric}
}

The current theoretical status has been summarized in the preceding section. 
The novel calculations of NNLO QCD \cite{Hermann:2013kca} and NLO EW 
\cite{Bobeth:2013tba} short-distance corrections to $C_{10}(\mu_b)$ remove
important renormalization scheme dependences at the matching scale present
at lower orders. The prediction of branching ratios is formally not
complete at the NLO in QED due to the lack of virtual corrections to the matrix
element of (\ref{eq:dim-6:eff:Lag}). These corrections will remove scheme and
the $\mu_b$ scale dependence still present in current predictions, however, they
do not undergo extra enhancement, contrary to the included ones that receive
either $1/s^2_W$ or $\ln^2(m_W/\mu_b)$ enhancement factors. In consequence,
the residual $\mu_b$ dependence of our result is very weak, as discussed in 
the previous section.

Based on (\ref{eq:LOQED:ME}) and (\ref{eq:BR}), most recent
predictions for the branching ratios in the SM have been presented \cite{Bobeth:2013uxa}
for all decay channels
\begin{equation}
  \label{eq:BR:prediction}
\begin{aligned}
  \overline{{\cal B}}[B_s \to \bar{e}e] & = (8.54 \pm 0.55) \times 10^{-14}\,, \qquad &
  \overline{{\cal B}}[B_d \to \bar{e}e] & = (2.48 \pm 0.21) \times 10^{-15}\,, 
\\  
  \overline{{\cal B}}[B_s \to \bar{\mu}\mu] & = (3.65 \pm 0.23) \times 10^{-9}\,, &
  \overline{{\cal B}}[B_d \to \bar{\mu}\mu] & = (1.06 \pm 0.09) \times 10^{-10}\,, 
\\
  \overline{{\cal B}}[B_s \to \bar{\tau}\tau] & = (7.73 \pm 0.49) \times 10^{-7}\,, &
  \overline{{\cal B}}[B_d \to \bar{\tau}\tau] & = (2.22 \pm 0.19) \times 10^{-8}\,, 
\end{aligned}
\end{equation}
with the specified input parameters in that work. The error budget is listed in table
\ref{tab:BR:errors}, showing that currently the largest uncertainties are caused
by the imprecise knowledge of CKM parameters and the decay constants. Adding up
in quadrature the various uncertainties gives rise to a total relative uncertainty
of less than $7\,\%$ for $B_s\to \bar\ell\ell$ and below $9\,\%$ for $B_d\to \bar\ell\ell$ 
decays. 

Concerning the CKM elements, latest results of 2013 fits from the CKMfitter \cite{Charles:2004jd}
and UTfit (post-EPS13) \cite{Ciuchini:2000de} groups have been employed for $|V_{tb}^{} V_{td}^*|$
and the ratio $|V_{tb}^{} V_{ts}^* / V_{cb}^{}|$, where the latter is to very
good approximation insensitive to $V_{cb}$. For the purpose of the predictions of
$B_s\to \bar\ell\ell$ modes, the value of $|V_{cb}|_{\, \rm incl} = 0.04242 \pm 0.00086$
has been used, which was determined from inclusive semi-leptonic $b\to c \ell \nu_\ell$
decays \cite{Gambino:2013rza}. This value differs by $3.0\,\sigma$ from 
$|V_{cb}|_{\, \rm excl} = 0.03904 \pm 0.00075$ as determined from exclusive 
$\bar{B}\to D^* \ell \bar{\nu}_\ell$ decays, based on recent lattice predictions
of $B\to D^*$ form factors \cite{Bailey:2014tva}. The usage of the exclusive
value of $V_{cb}$ instead of the inclusive would lower the branching ratios of 
$B_s$ decays by a sizeable amount  of $(|V_{cb}|_{\, \rm excl} / |V_{cb}|_{\, \rm incl})^2 
\sim 15\,\%$, i.e., for example $\overline{{\cal B}}[B_s \to \bar{\mu}\mu] = (3.09 \pm 0.19) 
\times 10^{-9}$. This result is by $2.4\,\sigma$ lower than the one obtained with
the inclusive-$V_{cb}$ value (\ref{eq:BR:prediction}), showing the importance 
of a clarification of the discrepancy of inclusive and exclusive determinations 
of $V_{cb}$. In this respect, one might note that the UTfit group
provides a global fit where $|V_{cb}|_{\,\rm excl} = 0.03955 \pm 0.00088$ 
(and $|V_{ub}|_{\,\rm excl}$) has been used as prior and the fitted posterior
value $|V_{cb}| = 0.04121 \pm 0.00050$ was obtained, indicating that other data
employed in the global fit prefers also larger values of $V_{cb}$.
 
Other uncertainties are below the $2\,\%$ level. A number of uncertainties 
related to higher order corrections are combined in the column labeled 
``non-parametric'' in table \ref{tab:BR:errors}. They comprise higher order
perturbative ${\cal O}(\alpha_s^3, \alpha_e^2, \alpha_s^{}\alpha_e^{})$ matching
corrections estimated in \cite{Hermann:2013kca,Bobeth:2013tba}, neglected
${\cal O}(\alpha_e)$ corrections to (\ref{eq:BR}), the truncated $\rm dim = 8$ 
contributions in the OPE of electroweak interactions (\ref{eq:eff:Lag}) and
others as specified in \cite{Bobeth:2013uxa}.

\begin{table}[t]
\caption[]{ 
  Relative uncertainties from various sources in $\overline{{\cal B}}[B_s\to \bar\ell\ell]$ 
  and $\overline{{\cal B}}[B_d\to \bar\ell\ell]$. In the last column they are added 
  in quadrature.}
\label{tab:BR:errors}
\vspace{0.4cm}
\begin{center}
\renewcommand{\arraystretch}{1.4}
\begin{tabular}{c||ccc|cc|c|c|c}
& $f_{B_q}$ 
& CKM 
& $\tau_H^q$ 
& $M_t$ 
& $\alpha_s$ 
& other param.     
& non-param.        
& $\sum$
\\
\hline \hline
  $\overline{\mathcal{B}}\big[B_s\to\bar\ell\ell\big]$
& $4.0 \,\%$
& $4.3 \,\%$ 
& $1.3 \,\%$
& $1.6 \,\%$
& $0.1 \,\%$
& $< 0.1 \,\%$
& $1.5 \,\%$
& $6.4 \,\%$
\\
  $\overline{\mathcal{B}}\big[B_d\to\bar\ell\ell\big]$
& $4.5 \,\%$
& $6.9 \,\%$
& $0.5 \,\%$
& $1.6 \,\%$
& $0.1 \,\%$
& $< 0.1 \,\%$
& $1.5 \,\%$
& $8.5 \,\%$
\end{tabular}
\renewcommand{\arraystretch}{1.0}
\end{center}
\end{table}

In view of the sizeable uncertainties due to CKM elements and decay constants 
one might consider in the SM the ratio of the branching ratio $B_q\to \bar\ell\ell$
and the mass difference of the neutral $B_q \overline{B}_q$ system, $\Delta M_q$,
\begin{align}
  \label{eq:def:kappa}
  \kappa_{q\ell} &
  \equiv \frac{\overline{{\cal B}}[B_q\to \bar\ell\ell] \, \Gamma_H^q \,(\Delta M_{B_q})^{-1}}
         {(G_F \, m_W \, m_\ell)^2 \sqrt{1 - 4 m_\ell^2/m_{B_q}^2}} 
  \stackrel{\rm SM}{=} \frac{3 |C_{10}(\mu_b)|^2}{4 \pi^3 \, C_{LL}(\mu_b) B_{B_q}(\mu_b)} \,.
\end{align}
In this ratio, both, CKM elements and decay constants, cancel and the only 
remaining nonperturbative quantity is the so-called bag factor $B_{B_q}$ \cite{Aoki:2013ldr}.
Here $C_{LL}$ denotes the Wilson coefficient of the SM $\Delta B = 2$ operator
(of $\rm dim = 6$). In the SM the theoretical prediction for the rhs 
of (\ref{eq:def:kappa}) yields \cite{Bobeth:2013uxa}
\begin{align}
  \label{eq:kappa:SM}
  \kappa_{s\ell} & = 0.0126 \pm 0.0007 \,, &
  \kappa_{d\ell} & = 0.0132 \pm 0.0012 \,, &  
\intertext{which is dominated by the uncertainty of the bag factors, whereas 
the lhs of (\ref{eq:def:kappa}) together with~(\ref{eq:exp:Br}) gives
the following experimental values}
  \kappa_{s\ell}|_{\rm exp.} & = 0.0104 \pm 0.0025 \,, &
  \kappa_{d\ell}|_{\rm exp.} & = 0.047  \pm 0.020  &  
\end{align}
that are consistent with the SM predictions. It can be seen that the overall
theory uncertainties in $\overline{{\cal B}}[B_q\to \bar\ell\ell]$ and
$\kappa_{q\ell}$ are quite similar at present.

Alternatively, one might use the experimental measurement of $\Delta M_q|_{\rm exp.}$
to determine the product $f_{B_q}^2 |V_{tb}^{} V_{tq}^\ast|^2$ under the
assumption of the SM and subsequently predict $\overline{{\cal B}}[B_q\to 
\bar\ell\ell]$, as proposed in \cite{Buras:2003td}, see \cite{Knegjens:2014}
for a recent update. Recasting (\ref{eq:def:kappa}) into
\begin{align}
  \overline{{\cal B}}[B_q\to \bar\ell\ell] &
  = \kappa_{q\ell} \, \frac{\Delta M_q|_{\rm exp.}}{\Gamma_H^q} \, 
    \, \sqrt{1- \frac{4 m_\ell^2}{m_{B_q}^2}} \, (G_F\, m_W\, m_\ell)^2
     \,,
\end{align}
it is straightforward to derive the theoretical uncertainties for the branching
ratio from the ones of $\kappa_{q\ell}$ and in addition also $\Gamma_H^q$,
whereas the ones of $\Delta M_q|_{\rm exp.}$ are numerically negligible
at the current level of precision. With the latest experimental
numbers $\Delta M_q|_{\rm exp.}$ and the SM predictions of $\kappa_{q\ell}$
(\ref{eq:kappa:SM}) one obtains
\begin{align}
  \overline{{\cal B}}[B_s\to \bar\mu\mu] & = (3.53 \pm 0.20) \times 10^{-9}\,, &
  \overline{{\cal B}}[B_d\to \bar\mu\mu] & = (1.00 \pm 0.09) \times 10^{-10}\,,
\end{align} 
which yields in the case of $B_s\to \bar\mu\mu$ values closer to the predictions
based on $|V_{cb}|_{\rm incl}$ than on $|V_{cb}|_{\rm excl}$.

%
%
%

\section{Conclusions
}

The latest standard model predictions have been presented for the branching ratios
of rare leptonic $B_q \to \bar\ell\ell$ ($q = d,s$ and $\ell = e,\mu,\tau$) decays.
The inclusion of NLO electroweak and NNLO QCD corrections decrease previous
renormalization scheme dependences from about $7\,\%$ to $0.6 \,\%$ and 
$1.8\,\%$ to $0.2 \,\%$, respectively. The current uncertainties of the branching
ratios are below $7\,\%$ for $B_s \to \bar\ell\ell$ and $9\,\%$ for $B_d \to 
\bar\ell\ell$ channels, where the dominant source are the imprecise knowledge
of the CKM matrix elements $V_{cb}$ and $V_{td}$ and comparable uncertainties
from the decay constants of $B_{q}$ mesons. In this respect, the ratios $\kappa_{q\ell}
\propto \overline{{\cal B}}[B_q\to \bar\ell\ell]/\Delta M_q$ are free of the decay
constant and CKM factors in the SM. Their current predictions are dominated by
the uncertainties due to the bag factors in $\Delta B = 2$ hadronic matrix 
elements and have comparable size to the ones of the branching ratios.

%
%
%

\section*{Acknowledgments}

I am indebted to the organizers of the Moriond EW 2014 conference for the opportunity
to present a talk and the kind hospitality. I thank Martin Gorbahn, Thomas 
Hermann, Miko{\l}aj Misiak, Emmanuel Stamou and Matthias Steinhauser for our
fruitful collaboration and comments on this manuscript, and Robert Knegjens for
useful discussions. This work received partial support from the ERC Advanced
Grant project ``FLAVOUR'' (267104).

%
%
%


\section*{References}

\begin{thebibliography}{99}

\bibitem{Aaij:2013aka}
  R.~Aaij {\it et al.}  [LHCb Collaboration],
  {\em Phys.\ Rev.\ Lett.}  {\bf 111}, 101805 (2013)
  [arXiv:1307.5024 [hep-ex]].

\bibitem{Chatrchyan:2013bka}
  S.~Chatrchyan {\it et al.}  [CMS Collaboration],
  {\em Phys.\ Rev.\ Lett.}  {\bf 111}, 101804 (2013)
  [arXiv:1307.5025 [hep-ex]].

\bibitem{DeBruyn:2012wk}
  K.~De Bruyn, R.~Fleischer, R.~Knegjens, P.~Koppenburg, M.~Merk, A.~Pellegrino and N.~Tuning,
  {\em Phys.\ Rev.\ Lett.} {\bf 109}, 041801 (2012)
  [arXiv:1204.1737 [hep-ph]].

\bibitem{LHCb-CMS:2013:EPSHEP}
  CMS and LHCb Collaborations,
  CMS-PAS-BPH-13-007, LHCb-CONF-2013-012, {\tt http://cds.cern.ch/record/1564324}.

\bibitem{Bediaga:2012py} 
  R.~Aaij {\it et al.}  [LHCb Collaboration],
  {\em Eur.\ Phys.\ J.} C {\bf 73}, 2373 (2013)
  [arXiv:1208.3355 [hep-ex]].

\bibitem{CMS:2013:prospects}
  CMS Collaboration,
  CMS-PAS-FTR-13-022, {\tt http://cds.cern.ch/record/1605250}.

\bibitem{Bobeth:2003at} 
  C.~Bobeth, P.~Gambino, M.~Gorbahn and U.~Haisch,
  {\em JHEP} {\bf 0404}, 071 (2004)
  [hep-ph/0312090].

\bibitem{Huber:2005ig} 
  T.~Huber, E.~Lunghi, M.~Misiak and D.~Wyler,
  {\em Nucl.\ Phys.} B {\bf 740}, 105 (2006)
  [hep-ph/0512066].

\bibitem{Misiak:2011bf} 
  M.~Misiak,
  arXiv:1112.5978 [hep-ph].

\bibitem{Inami:1980fz} 
  T.~Inami and C.~S.~Lim,
  {\em Prog.\ Theor.\ Phys.}  {\bf 65}, 297 (1981)
  [{\em Erratum-ibid.}  {\bf 65}, 1772 (1981)].

\bibitem{Buchalla:1992zm} 
  G.~Buchalla and A.~J.~Buras,
  {\em Nucl.\ Phys.} B {\bf 398}, 285 (1993).

\bibitem{Buchalla:1993bv} 
  G.~Buchalla and A.~J.~Buras,
  {\em Nucl.\ Phys.} B {\bf 400}, 225 (1993).

\bibitem{Misiak:1999yg} 
  M.~Misiak and J.~Urban,
  {\em Phys.\ Lett.} B {\bf 451}, 161 (1999)
  [hep-ph/9901278].

\bibitem{Buchalla:1998ba} 
  G.~Buchalla and A.~J.~Buras,
  {\em Nucl.\ Phys.} B {\bf 548}, 309 (1999)
  [hep-ph/9901288].

\bibitem{Hermann:2013kca}
  T.~Hermann, M.~Misiak and M.~Steinhauser,
  {\em JHEP} {\bf 1312}, 097 (2013)
  [arXiv:1311.1347 [hep-ph]].

\bibitem{Buchalla:1997kz} 
  G.~Buchalla and A.~J.~Buras,
  {\em Phys.\ Rev.} D {\bf 57}, 216 (1998)
  [hep-ph/9707243].

\bibitem{Buras:2012ru} 
  A.~J.~Buras, J.~Girrbach, D.~Guadagnoli and G.~Isidori,
  {\em Eur.\ Phys.\ J.} C {\bf 72}, 2172 (2012)
  [arXiv:1208.0934 [hep-ph]].

\bibitem{Bobeth:2013tba}
  C.~Bobeth, M.~Gorbahn and E.~Stamou,
  {\em Phys.\ Rev.} D {\bf 89}, 034023 (2014)
  [arXiv:1311.1348 [hep-ph]].

\bibitem{Brod:2010hi} 
  J.~Brod, M.~Gorbahn and E.~Stamou,
  Phys.\ Rev.\ D {\bf 83}, 034030 (2011)
  [arXiv:1009.0947 [hep-ph]].

\bibitem{Aoki:2013ldr} 
  S.~Aoki, Y.~Aoki, C.~Bernard, T.~Blum, G.~Colangelo, M.~Della Morte, S.~D{\"u}rr and A.~X.~E.~Khadra {\it et al.},
  arXiv:1310.8555 [hep-lat].

\bibitem{Aditya:2012im} 
  Y.~G.~Aditya, K.~J.~Healey and A.~A.~Petrov,
  {\em Phys.\ Rev.} D {\bf 87}, 074028 (2013)
  [arXiv:1212.4166 [hep-ph]].

\bibitem{Golonka:2005pn} 
  P.~Golonka and Z.~Was,
  {\em Eur.\ Phys.\ J.} C {\bf 45}, 97 (2006)
  [hep-ph/0506026].

\bibitem{Bobeth:2013uxa}
  C.~Bobeth, M.~Gorbahn, T.~Hermann, M.~Misiak, E.~Stamou and M.~Steinhauser,
  {\em Phys.\ Rev.\ Lett.}  {\bf 112}, 101801 (2014) 
  [arXiv:1311.0903 [hep-ph]].

\bibitem{Charles:2004jd} 
  J.~Charles {\it et al.}  [CKMfitter Group Collaboration],
  {\em Eur.\ Phys.\ J.} C {\bf 41}, 1 (2005)
  [hep-ph/0406184], updates at {\tt http://ckmfitter.in2p3.fr}.

\bibitem{Ciuchini:2000de} 
  M.~Ciuchini, G.~D'Agostini, E.~Franco, V.~Lubicz, G.~Martinelli, F.~Parodi, P.~Roudeau and A.~Stocchi,
  {\em JHEP} {\bf 0107}, 013 (2001)
  [hep-ph/0012308], updates at {\tt http://www.utfit.org}.

\bibitem{Gambino:2013rza} 
  P.~Gambino and C.~Schwanda,
  {\em Phys.\ Rev.} D {\bf 89}, 014022 (2014)
  [arXiv:1307.4551 [hep-ph]].

\bibitem{Bailey:2014tva} 
  J.~A.~Bailey, A.~Bazavov, C.~Bernard, C.~M.~Bouchard, C.~DeTar, D.~Du, A.~X.~El-Khadra and J.~Foley {\it et al.},
  arXiv:1403.0635 [hep-lat].

\bibitem{Buras:2003td} 
  A.~J.~Buras,
  {\em Phys.\ Lett.} B {\bf 566}, 115 (2003)
  [hep-ph/0303060].

\bibitem{Knegjens:2014} 
  R.~Knegjens, talk given at DIS 2014, 28 April to 2 May, 2014, Warsaw, Poland,
  {\tt http://indico.cern.ch/event/258017/session/5/contribution/255}.

\end{thebibliography}
\end{document}